\documentclass[%
aps,prl,
 amsmath,amssymb,
 reprint,%
 superscriptaddress,
]{revtex4-1}

\usepackage{graphicx}
\usepackage{dcolumn}
\usepackage{bm}
\usepackage{xcolor}
\usepackage{mathrsfs}

\usepackage[colorlinks,citecolor=magenta]{hyperref}

\graphicspath{{./Figures/}}

\begin{document}

\title{Percolation transition of pusher-type microswimmers}

\author{Fabian Jan Schwarzendahl}
\affiliation{Max Planck Institute for Dynamics and Self-Organization, Am Fa{\ss}berg 17, 37077 G\"ottingen, Germany}
\author{Marco G. Mazza}
\affiliation{Interdisciplinary Centre for Mathematical Modelling and Department of Mathematical Sciences, Loughborough University, Loughborough, Leicestershire LE11 3TU, United Kingdom}
\affiliation{Max Planck Institute for Dynamics and Self-Organization, Am Fa{\ss}berg 17, 37077 G\"ottingen, Germany}

\date{\today}

\begin{abstract}
We identify the presence of a continuum percolation transition in model suspensions of pusher-type microswimmers. 
The clusters dynamically aggregate and disaggregate resulting from a competition of attractive and repulsive hydrodynamic and steric interactions. As the microswimmers' filling fraction increases, the cluster size distribution approaches a scale-free form and there emerge large clusters spanning the entire system. We characterize this microswimmer percolation transition via the critical exponents associated to cluster size distribution $\tau$, correlation length $\nu$, mean cluster size $\gamma$, and clusters' fractal dimension $d_f$. We are able to show that two scaling relations, known from percolation theory, also hold for our microswimmers.  A real-space renormalization group transformation can approximately predict the value of the exponent $\nu$.  This finding opens new vistas on microswimmers' congregative processes. 
\end{abstract}

\maketitle

The spontaneous organization of suspensions of  microswimmers such as bacteria or microalgae is a fascinating result of the concurrence of physical and biological forces~\cite{elgetiRepPRogPhys2015}. 
Extensive investigations have shown numerous intriguing effects such as
self-concentration~\cite{SchwarzendahlSM2018,dombrowskiPRL2004},
swarming~\cite{copelandSM2009}, 
spontaneous formation of spiral vortices~\cite{wiolandPRL2013}, spontaneous formation of fluid flows~\cite{AtisPRX2019}, or
bacterial turbulence~\cite{wensinkPNAS2012}. These phenomena arise from the arena of nonequilibrium physics, where energy stored from different sources is converted in active motion. 

The study of clustering or self-concentration is especially stimulating, as this has important implications to the formation of biofilms.
Active Brownian particles (ABPs) show a clustering effect termed motility induced phase separation (MIPS) at large enough filling fractions \cite{ButtinoniPRL2013,FilyPRL2012,RednerPRL2013,BialkeEPL2013,StenhammerPRL2013,WysockiEPL2014}, 
and the nature of this transition to the clustering state has been characterized \cite{digregorioPRL2018}.
Studies of squirmers have however shown that MIPS is suppressed if hydrodynamic interactions are included \cite{Matas-NavarroPRE2014,TheersSM2018,ZottlPRL2014,AlarconSM2017}.
MIPS is distinct from the density heterogeneities found in systems of microswimmers 
\cite{SchwarzendahlSM2018,ishikawajfm2008}, 
which are mediated by hydrodynamic interactions, and occur at considerably lower filling fractions, as seen in particle approaches \cite{SchwarzendahlSM2018,ishikawajfm2008}, and also in continuum approaches~\cite{SchwarzendahlSM2018,BaskaranPNAS2009,EzhilanPoF2013,SaintillanPRL2008,UnderhillPRL2008}.

Another route to aggregation could be the percolation transition. For example, C-shaped ABPs without hydrodynamic interactions have been found to exhibit a percolation transition~\cite{hoellJCP2016}. Very recent experiments~\cite{MathijssenNature2019} have shown that percolating clusters of \textit{Spirostomum ambiguum} communicate through hydrodynamic trigger waves. Furthermore, studying the onset of mesoscale turbulence revealed a connection to directed percolation \cite{doostmohammadi2017onset}. Thus, it is necessary to include hydrodynamic interaction to model realistic conditions.

Here, we show that, upon increasing a suspension's filling fraction, model biological microswimmers undergo a percolation transition. 
We find that classical tools from percolation theory \cite{stauffer2014introduction} can be used to characterize the transition also by means of their critical exponents.

\begin{figure}
        \centering
        \includegraphics[width=0.7\columnwidth]{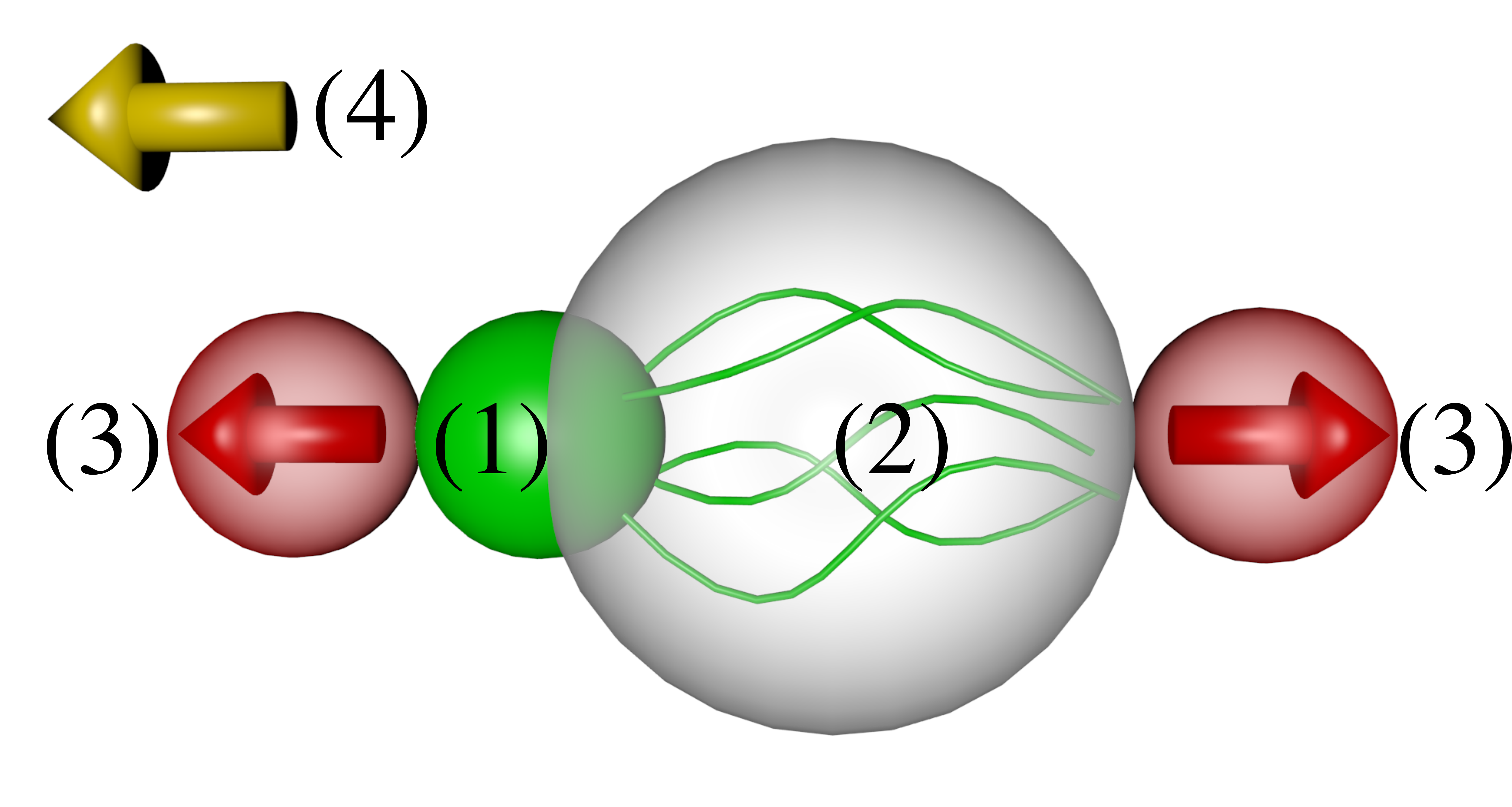}
        \caption{Schematic of our stroke-averaged pusher-type microswimmer. The green sphere (1) models the swimmers body and the transparent sphere (2) the stroke-average of the flagellar beat. The red spheres with arrows (3) mark the location of the regularized force poles. The golden arrow (4) represents the swimming direction.
}
        \label{fig:SwimmerModel}
\end{figure}

To simulate the collective behavior of microswimmers, such as the pusher-type bacteria \textit{Escherichia coli}
we use a stroke-averaged biological microswimmer model \cite{SchwarzendahlSM2018}. 
A schematic of the stroke averaged microswimmer is shown in Fig.~\ref{fig:SwimmerModel}. 
The dynamics of this dumbbell are governed by Newton's equations of motion~\cite{SchwarzendahlSM2018}. 
As the swimmer is an asymmetric dumbbell, the hydrodynamic center is shifted away from the center of mass of the swimmer. This symmetry breaking enables the propulsion of the swimmer into the direction of the golden arrow (4) in Fig.~\ref{fig:SwimmerModel}. 

\begin{figure*}
        \centering
        \includegraphics[width=1.8\columnwidth]{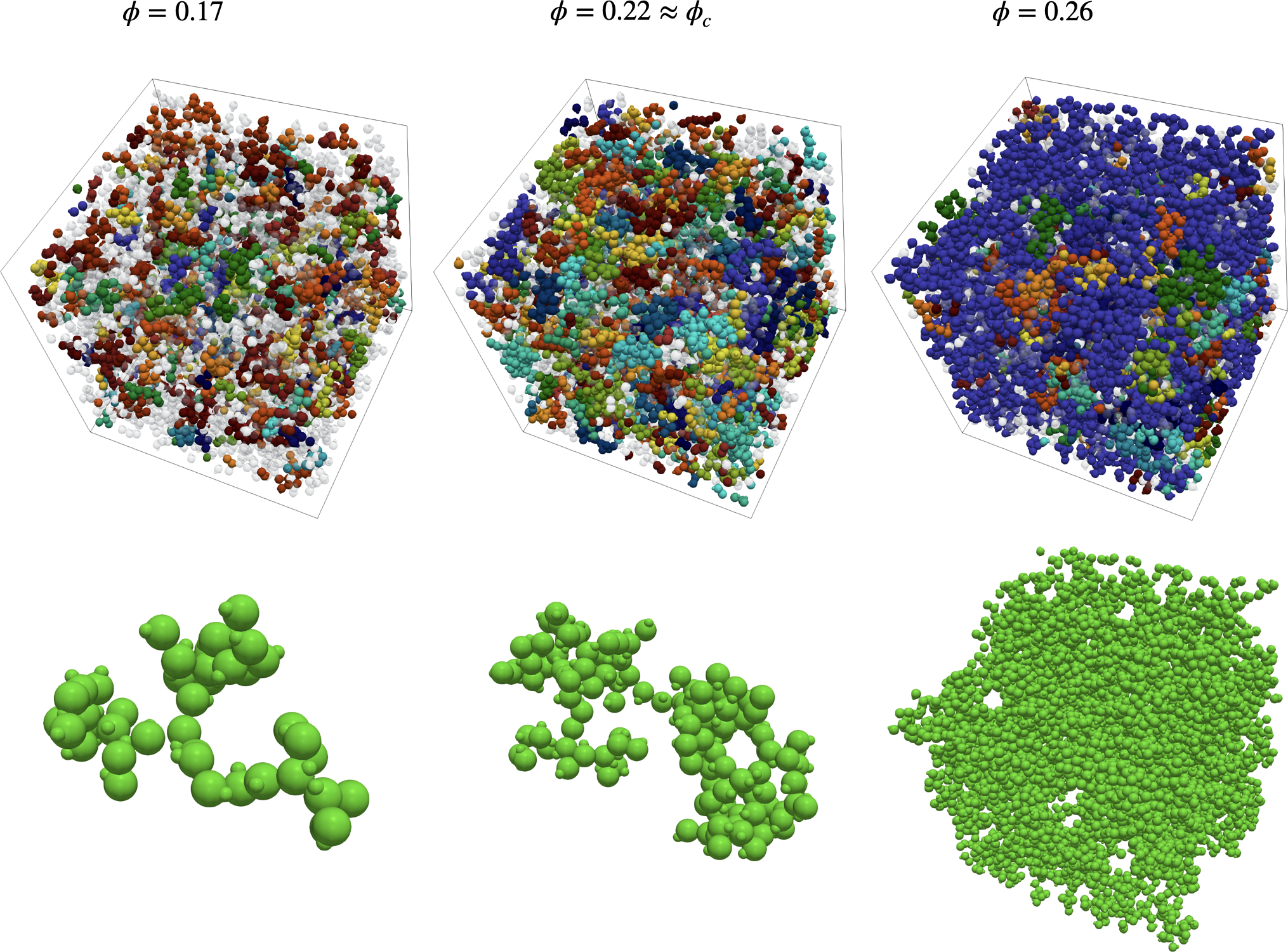}
        \caption{Top row: example configurations of swimmers at different filling fractions (large system). 
        Gray transparent particles do not belong to any cluster. Different colors correspond to different clusters. The system spanning cluster at $\phi=0.26$ indicates a percolation transition. Bottom row: zoom-in view of the largest cluster identified in the corresponding configuration in the top row.}
        \label{fig:configExamples}
\end{figure*}

In addition to the steric interactions between the microswimmers, we take into account the hydrodynamic flow field that is generated by each microswimmer, since these interactions play a key role in the collective behavior. 
Experiments~\cite{DrescherPNAS2011} have shown that the hydrodynamic flow field of the pusher-type swimmer \textit{Escherichia coli} corresponds to a force dipole.
Accordingly we account for the flow field by adding regularized force regions, which are shown as red spheres (3) in Fig.~\ref{fig:SwimmerModel}.
Furthermore, no-slip boundary conditions are implemented on the dumbbell-shaped swimmer's body.

To resolve the hydrodynamic interaction numerically, we use the multiparticle  collision  dynamics (MPCD) technique. 
MPCD is a particle based method accounting for hydrodynamic modes up to the Navier--Stokes level \cite{MalevanetsJCP1999,gompperBookChap2009}. 
Specifically, we use the MPCD-at+A\cite{gompperBookChap2009,NoguchiEPL2007,GotzePRE2007} technique, that conserves the fluids' temperature. 
The MPCD fluid is characterized by its temperature $T$, MPCD particle mass $m$, and size of a MPCD grid cell $a$ as a unit of length. 
Our simulations are performed with an average of $\langle N_\mathsf{C} \rangle = 20$ MPCD particles in each MPCD cell; we carry out simulations with a timestep $\delta t = 0.01$. 
The size of an individual swimmer is $\sigma \approx 5a$. 
We use a cubic domain with periodic boundary conditions and constant volume. 

We perform simulations of both a small system of linear size $L=100a$, and a large system $L=200a$.
In the small system we simulate a range of $N=388-2300$ swimmers, corresponding to filling fractions from $\phi = 0.07$ to $\phi= 0.43$, while in the large system we use $N=3800-18200$, corresponding to filling fractions from $\phi = 0.09$ to $\phi= 0.43$. 
The active state is characterized by the P\'eclet number $\mathcal{P}=  { v \sigma}/{D}$, which compares the self-propulsion speed $v$ (active transport) to the diffusive transport $D$ of a microswimmer.
Additionally the flow around our swimmers is characterized by the Reynolds number (measuring the ratio of inertial to viscous forces) $\mathcal{R}=\sigma v \rho/\eta$, where $\rho$ is the fluids' density and $\eta$ is the fluids' viscosity. 
See \cite{SchwarzendahlSM2018} for more computational details.
In all our simulations the P\'eclet number is $\mathcal{P} \approx 2.6 \times 10^3$ and the Reynolds number is $\mathcal{R}=0.1$.


Figure~\ref{fig:configExamples} (top row) shows typical, steady-state configurations of  our active pusher-type swimmers at three different filling fractions $\phi$. Clusters of particles are temporarily held together by attractive regions in the swimmers' pair distribution function, due to the interplay of hydrodynamic and steric effects~\cite{schwarzendahlJCP2019}.
We perform a cluster analysis based on the interparticle distance, \emph{i.e.}, if their center of mass distance is less than $r_\mathrm{cluster}= 1.5 \sigma$  particles are assigned to the same cluster.
Figure~\ref{fig:configExamples} shows the resulting clusters for those configurations in different colors, where the transparent particles belong to clusters with four or fewer particles. 
As the filling fraction increases clusters grow in size and at $\phi=0.22$ we find clusters comparable to the system size. 
For $\phi\geq 0.26$, very large clusters, spanning the entire system emerge. 
Figure~\ref{fig:configExamples} (bottom row) shows the largest cluster found in the corresponding system in the top row. Visual inspection of their three-dimensional structure reveals a planar-like conformation for $\phi\lesssim 0.22$ (see below for details about the fractal dimension). For $\phi=0.26$, a cluster spanning the entire domain is clearly visible. 


\begin{figure*}
        \centering
        \includegraphics[width=2.0\columnwidth]{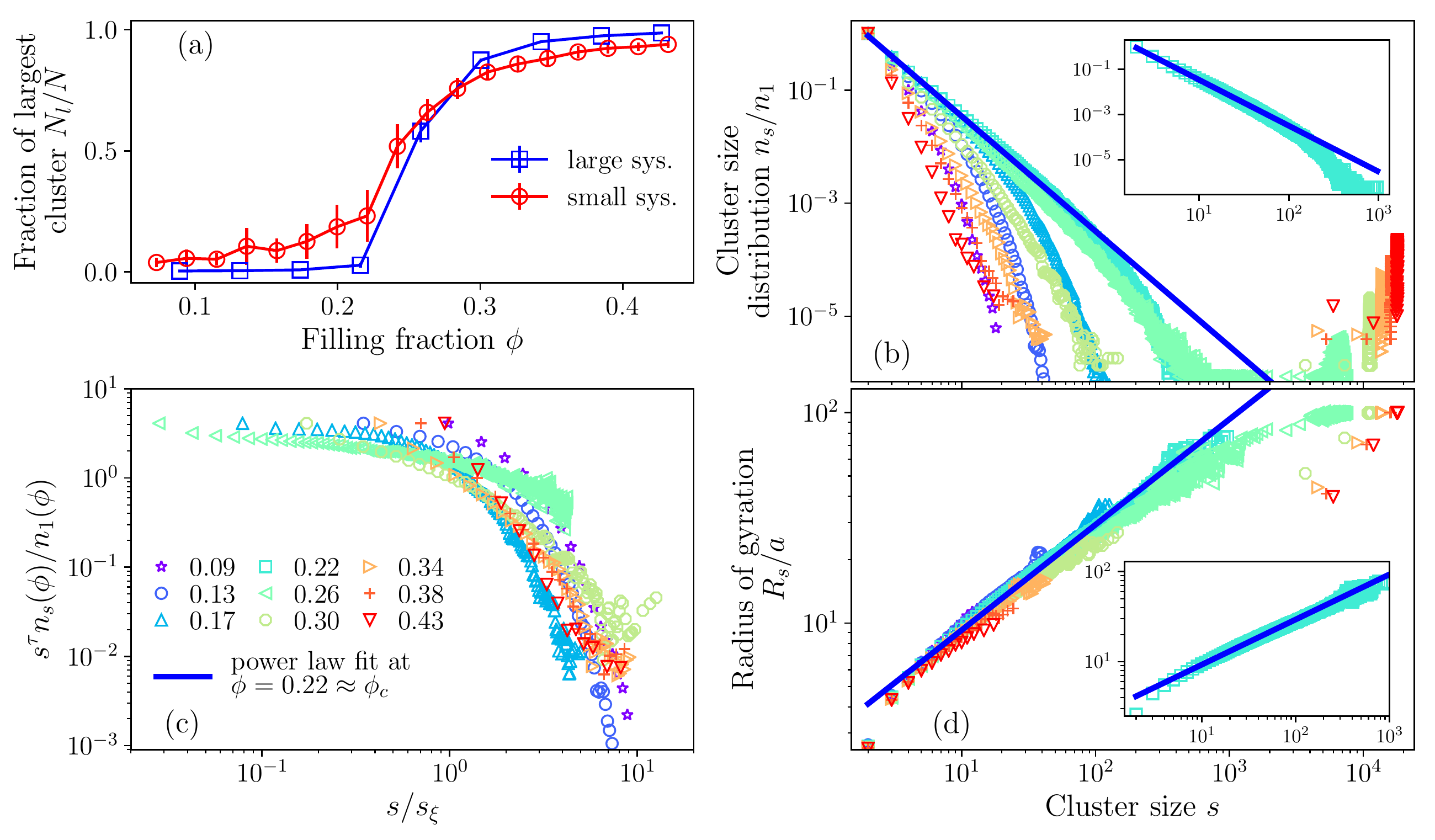}
        \caption{(a) Fraction of number of particles  in the largest cluster to total number of particles $N_l/N$. (b) Cluster size distribution $n_s$ normalized to the number of singletons $n_1$ for different filling fractions (large system). 
        The solid line is a fit to the data for $\phi=0.22$, 
        giving the Fisher exponent $\tau = 2.033 \pm 0.003$. 
        The inset (b) shows the fitted data ($\phi=0.22$) for the sake of clarity.
         (c) Collapse of the probability distribution to an exponential scaling function.
         (d) Radius of gyration for different filling fractions (large system). 
        The solid line is a fit to the data for $\phi=0.22$, 
        giving the fractal dimension $d_f = 1.99 \pm 0.02$. 
        The inset shows the fitted data ($\phi=0.22$) for the sake of clarity.
      }
        \label{fig:distributionplot}
\end{figure*}

To scrutinize the statistical properties of the active clusters, we compute the fraction of the largest cluster defined as the ratio of the number of particles in the largest cluster $N_l$ to the total number of particles in the system $N$.
Figure~\ref{fig:distributionplot}(a) shows the fraction of the largest cluster $N_l/N$ for varying filling fraction $\phi$ in the large as well as the small system. It can be seen that $N_l/N$ acts as an order parameter for the percolation transition, occurring at $\phi \approx 0.22$. 

Furthermore, we compute the cluster size distribution $n_s(s)$, where $s$ is the cluster size, \emph{i.e.}, the number of member particles. 
Figure~\ref{fig:distributionplot}(b) shows $n_s$ for different $\phi$ (data from our large scale simulations). 
As $\phi$ increases towards $\phi=0.22$, the cluster size distribution $n_s(s)$ increasingly approaches a scale-free form.
This cluster analysis of the swimmers' configurations points at a percolation transition, where the filling fraction takes the role of the occupation probability (as is common in continuum percolation of colloids~\cite{safranPRA1985} or simple fluids~\cite{heyesMolPhys1989}). 

At the percolation transition, it is expected that $n_s(s)\sim s^{-\tau}$, where $\tau$ is the Fisher exponent~\cite{stauffer2014introduction}. 
Identifying $\phi_c=0.22$ as the critical filling fraction (see Fig.~\ref{fig:distributionplot}(b) inset) we find the Fisher exponent $\tau \approx 2.033 \pm 0.003$.

Away from the percolation transition at $\phi_c$, the theory predicts that 
\begin{align}
n_s(s)\sim s^{-\tau} \mathrm{exp}\left(-\frac{s}{s_\xi}\right),
    \label{eq:generalnsrealtion}
\end{align}
where $s_\xi$ is the cutoff cluster size. We find $s_\xi$ for each filling fraction by fitting the respective cluster size distribution in Fig.~\ref{fig:distributionplot}(b) to Eq.\eqref{eq:generalnsrealtion}. 
In Fig.~\ref{fig:distributionplot}(c) we plot $ s^{\tau} n_s(\phi)/n_1(\phi)$ against $s/s_{\xi}$ and find collapse of the data onto an exponential master curve, as predicted by Eq.\eqref{eq:generalnsrealtion}. 





The geometrical features of the clusters can be described by their radius of gyration and fractal dimension. In Fig.~\ref{fig:distributionplot}(d) we show the dependence of the radius of gyration 
\begin{align}
 R_s^2\equiv  \frac{1}{N_{clust}} \sum_i(\bm{r_i}-\bm{r}_{\mathrm{CoM}})^2, 
\end{align}
on cluster size $s$ for different filling fractions (data from our large scale simulations). 
The distribution at filling fraction $\phi=0.22$ approaches a power law, as expected at percolation.
The resulting exponent is the fractal dimension $d_f = 1.99 \pm 0.02$. 
A fractal dimension of $d_f \approx 2$ indicates that our clusters are typically two dimensional, which can also be observed in the configurations depicted in Fig.~\ref{fig:configExamples}.

\begin{figure}
        \centering
            \includegraphics[width=0.95\columnwidth]{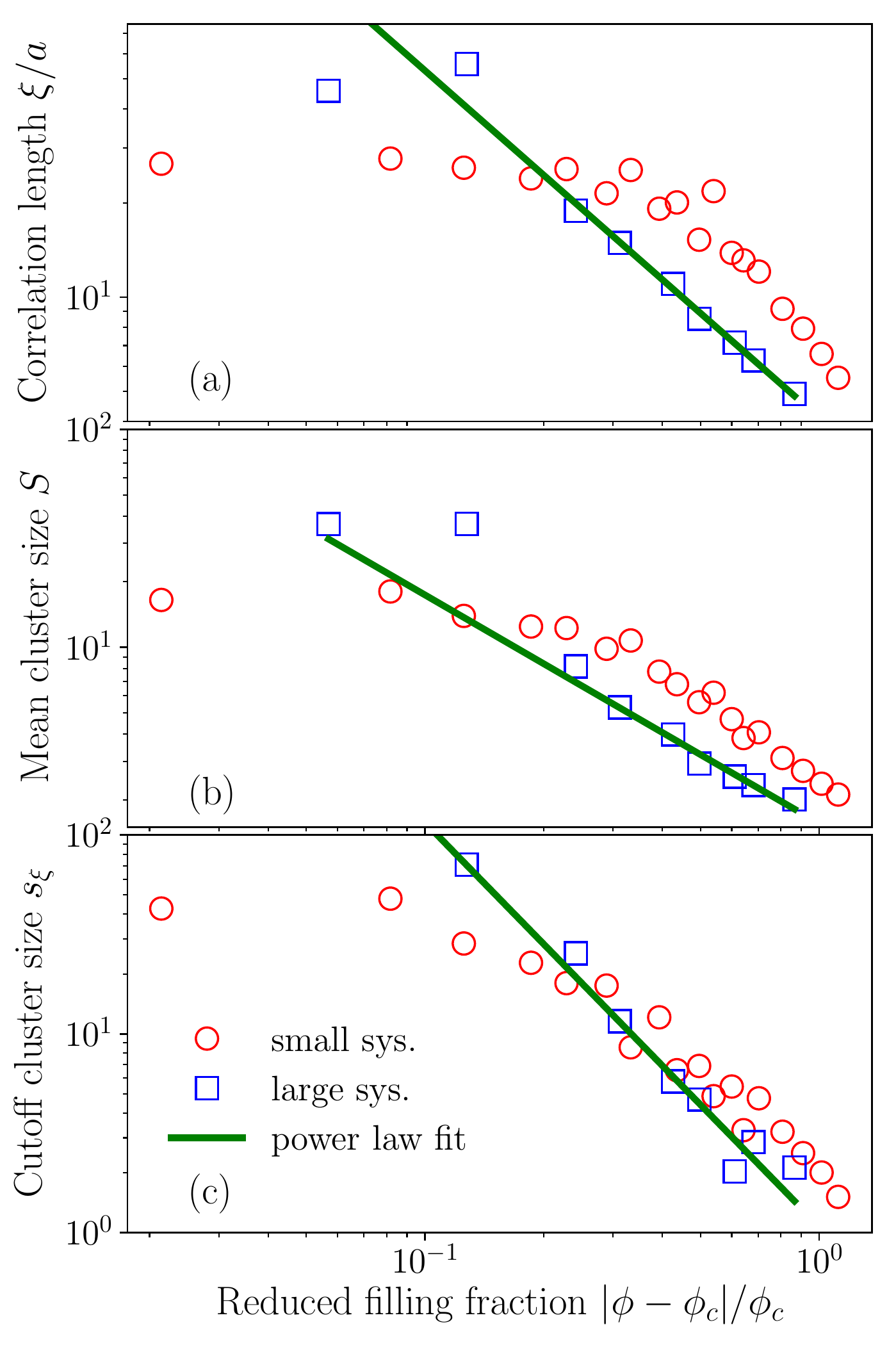}
        \caption{Dependence of the (a) correlation length, (b) mean cluster size, and (c) cutoff cluster size on the filling fraction $\phi$. 
        Solid green lines are power law fits to the data for the large systems. 
        For (a) the correlation length we find the critical exponent $\nu=1.11\pm 0.04$,
        for (b) the mean cluster size we find the critical exponent $\gamma= 1.05 \pm 0.08$ and for (c) the cutoff cluster size we find the critical exponent $\sigma= 0.49 \pm 0.09$.}
        \label{fig:sizecorrelation}
\end{figure}

Information on the radius of gyration $R_s$, allows us to compute the correlation length
\begin{align}
    \xi^2 = \frac{ 2 \sum_s R_s n_s s^2}{\sum_s n_s s^2}.
    \label{eq:CorrelationLength}
\end{align}
Percolation theory \cite{stauffer2014introduction} predicts a scaling behavior close to criticality $\phi_c$  
\begin{align}
    \xi \sim |\phi -\phi_c|^{-\nu}.
    \label{eq:CorrelationLengthScaling}
\end{align}

We now present a real-space renormalization group (RSRG) calculation of our percolation problem. We discretize the system by choosing a cubic lattice. The RSRG approximation consists in taking sites independently occupied with probability $p$. Next, we define a cubic cell that will play the role of the renormalized sites. We must choose a cell occupation probability $p'=\mathscr{R}(p)$ such that $\mathscr{R}(p)$ contains the essential physics of our percolation problem~\cite{reynoldsJPhysC1977}. Our RSRG rescales the lattice spacing by a factor $b=2$ in each spatial dimension. We define a cell as occupied if and only if it contains a set of sites such that the cell percolates by means of planar clusters. 
The correlation length will be rescaled as $\xi'=b^{-1}\xi$, and the correlation length exponent can be found as 
\begin{equation}\label{eq:RG_nu}
    \nu=\frac{\ln b}{\ln \lambda}\,,
\end{equation}
where $\lambda$ is the eigenvalue of the RSRG transformation linearized around the fixed point $p_c$, \begin{equation}\label{eq:RG_lambda}
    \lambda\equiv \frac{d\mathscr{R}}{dp}(p)\biggr\rvert_{p=p_c}\,. 
\end{equation}
For our cubic lattice, a cell is occupied if a planar configuration of sites is occupied, that is if four sites arranged in a plane (along the Cartesian directions or the diagonals), and also when three sites are occupied. Thus, 
\begin{equation}\label{eq:RG_prop}
    p'=\mathscr{R}(p)=18p^4(1-p)^4+72p^3(1-p)^5\,.
\end{equation}
We identify $p_c$ with $\phi_c=0.229$, and by using Eq.~\eqref{eq:RG_nu}-\eqref{eq:RG_prop} we find $\nu\simeq 1.21$. We will see below that our RSRG estimate comes relatively close to 
the simulational evaluation of $\nu$.

Figure~\ref{fig:sizecorrelation}(a) shows the correlation length (Eq.~\eqref{eq:CorrelationLength}) for varying $|\phi -\phi_c|/\phi_c$.
From a best fit we obtain $\phi_c = 0.229$ (large system) and $\phi_c = 0.204$ (small system), where the  data exhibit power-law behavior. The percolation threshold $\phi_c$ is expected to be affected by finite-size effects.
We find a correlation length exponent, predicted in Eq.~\eqref{eq:CorrelationLengthScaling},    $\nu=1.11\pm 0.04$. 

The mean cluster size is defined as
\begin{align}
    S= \frac{\sum_s n_s s^2}{\sum_s n_s s},
    \label{eq:MeanClusterSize}
\end{align}
and percolation theory predicts that, close to criticality $\phi_c$, the mean cluster size scales as \cite{stauffer2014introduction}
\begin{align}
    S \sim |\phi -\phi_c|^{-\gamma} \ ,
    \label{eq:MeanClusterSizeScaling}
\end{align}
which defines the exponent $\gamma$.
Figure~\ref{fig:sizecorrelation}(b) shows the mean cluster size as a function of $|\phi -\phi_c|/\phi_c$, 
Similar to the correlation length (Fig.~\ref{fig:sizecorrelation}(a)), we obtained an optimal critical filling fractions  $\phi_c = 0.229$ (large system) and $\phi_c = 0.204$ (small system). 
Additionally, we fitted a power law distribution to test the predicted scaling law Eq.\eqref{eq:MeanClusterSizeScaling} and find  $\gamma= 1.05 \pm 0.08$.

Because our simulations are based on finite systems, finite-size scaling predicts that the growth of the mean cluster size $S$ is capped once the correlation length $\xi\approx L$, and that it obeys the general scaling $S(\xi,L)=\xi^{\gamma/\nu}s_1(L/\xi)$. 
Close to $\phi_c$, $\xi\gg L$ and we expect $S(\xi,L)\propto L^{\gamma/\nu}$.
This is verified by the  asymptotic values in Fig.~\ref{fig:sizecorrelation}(b). Again, these results bolster our assumption that the transition we find corresponds to a percolation transition.

The cutoff cluster size $s_{\xi}$ is also predicted to diverge  as the percolation threshold is approached 
\begin{align}
    s_{\xi}\sim |\phi -\phi_c|^{- \frac{1}{\sigma}}\,,
    \label{eq:cutoffClusterScaling}
\end{align}
which defines the exponent $\sigma$.
Figure~\ref{fig:sizecorrelation}(c) shows the cutoff cluster size for varying reduced filling fraction $|\phi -\phi_c|/\phi_c$. Again, we obtain optimal critical filling fractions $\phi_c = 0.229$ for the large system and $\phi_c = 0.204$ for the small system. Furthermore, we show a fit to Eq.\eqref{eq:cutoffClusterScaling}, and find the critical exponent $\sigma= 0.49 \pm 0.09$.

Using the critical exponents and fractal dimension that we have found, we can now apply a more stringent test from percolation theory consisting in the following scaling relations~\cite{stauffer2014introduction}
\begin{align}
\sigma&= \frac{1}{\nu d_f},
    \label{eq:sigmarelation}
\\
\gamma&= \nu ( 2d_f -d),
    \label{eq:gammarelation}
\end{align}
where $d=3$ is the dimensionality of the system. In fact, the exponent  
computed from our simulations satisfy  Eqs.~\eqref{eq:sigmarelation}-\eqref{eq:gammarelation}. 
The validity of the scaling relations in Eqs.~\eqref{eq:sigmarelation}-\eqref{eq:gammarelation} provides the most compelling test for the presence of a percolation transition in our microswimmer system.

In conclusion, we have shown that pusher-type microswimmers exhibit a percolation transition $\phi_c=0.229$, with a probability distribution  approaching a scale-free form. For larger $\phi$, system-spanning clusters arise. Correlation length as well as mean cluster sizes follow scaling laws. We verified two classical scaling relations from percolation theory. This percolation mechanism might represent a congregative dynamics important in the transition from the planktonic state to biofilms. 

Finally, the critical exponents found in our microswimmer model do not correspond to any known percolation universality class. It is thus premature to speak of any universality class associated to microswimmers until more systems are investigated.

We gratefully acknowledge support from the Deutsche Forschungsgemeinschaft (SFB 937, project A20) and from the Max Planck Society. F.J.S. thanks the ICMM group at Loughbourough University for the kind hospitality which enabled this work.


%

\end{document}